\documentclass[aps,prb,amsmath,amssymb,twocolumn,showpacs]{revtex4-1}

\usepackage{graphicx}
\usepackage{dcolumn}
\usepackage{bm}
\usepackage{color}

\begin{document}

\title{Extraordinary transverse magneto-optical Kerr effect in a superlens}

\author{E. Moncada-Villa$^{1,2}$}
\author{A. Garc\'{\i}a-Mart\'{\i}n$^{3}$}
\author{J. C. Cuevas$^2$}
\email{juancarlos.cuevas@uam.es}

\affiliation{$^{1}$Departamento de F\'{\i}sica, Universidad del Valle, AA 25360, Cali, Colombia}

\affiliation{$^2$Departamento de F\'{\i}sica Te\'orica de la Materia Condensada
and Condensed Matter Physics Center (IFIMAC), Universidad Aut\'onoma de Madrid,
E-28049 Madrid, Spain}

\affiliation{$^{3}$IMM-Instituto de Microelectr\'onica de Madrid (CNM-CSIC), Isaac Newton 8, 
PTM, Tres Cantos, E-28760 Madrid, Spain}

\date{\today}

\begin{abstract}
It has been shown that a slab of a negative index material can behave as a superlens 
enhancing the imaging resolution beyond the wavelength limit. We show here that if 
such a slab possesses in addition some magneto-optical activity, it could act as an ideal 
optical filter and exhibit an extraordinary transverse magneto-optical Kerr effect.
Moreover, we show that losses, which spoil the imaging resolution of these lenses, are
a necessary ingredient to observe this effect.
\end{abstract} 


% 78.20.Ls = Magneto-optical effects
% 78.20.Ci = Optical constants (including refractive index, complex dielectric constant, absorption, 
%            reflection and transmission coefficients, emissivity)
% 78.66.Bz = Metals and metallic alloys
% 42.25.Gy = Edge and boundary effects; reflection and refraction

\maketitle

As Veselago predicted in his seminal work,\cite{Veselago1968} negative index materials 
(NIMs) can exhibit very striking electromagnetic properties such as a reversed Doppler effect
or negative refraction. This latter property was exploited by Pendry to propose that
a slab of a NIM could behave as a superlens,\cite{Pendry2000} \emph{i.e.}\ it could allow the 
imaging of objects with subwavelength precision. Although the ideal realization of a superlens
in the optical regime is quite challenging, the basic ideas behind this concept have been 
confirmed\cite{Shalaev2007} and Pendry's proposal served as an inspiration that led to the
blossom of the field of metamaterials.\cite{Shalaev2007,Cai2009,Hess2012} These artificial
materials have had an impact in different subfields of optics. Thus for instance, chiral 
metamaterials, an alternative route to negative refraction, have been shown to boost 
the magnitude of optical effects such as optical activity, circular dichroism, or Faraday 
rotation.\cite{Mackay2004,Zhang2007,Zhu2014,Wang2009} The goal of this work is to show that 
NIMs and, in particular, superlenses also offer new fascinating possibilities for the field 
of magneto-optics.\cite{Zvezdin1997}

\begin{figure}[b]
\includegraphics[width=0.85\columnwidth,clip]{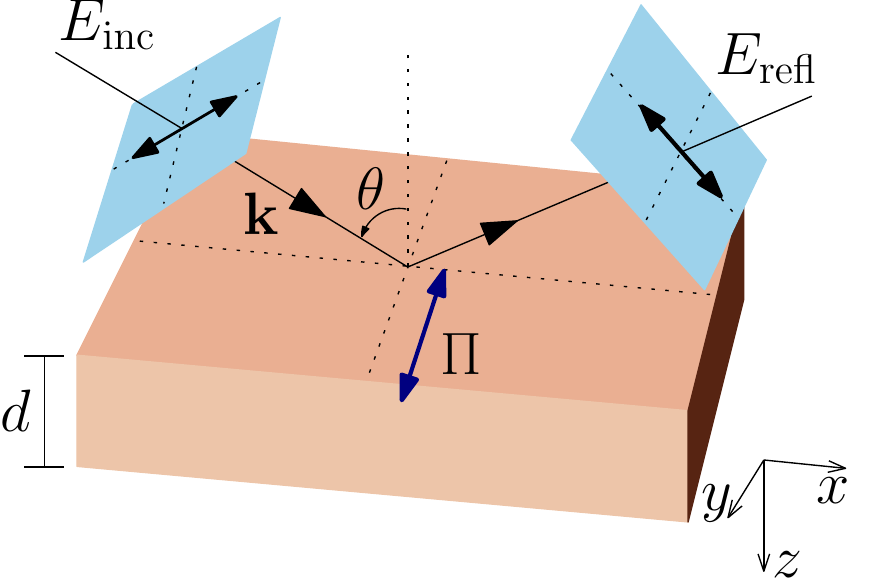}
\caption{(color online) Schematic representation of the transverse magneto-optical Kerr effect in a
slab, eventually made of a negative index material, with magneto-optical activity. In this
effect a $p$-polarized plane wave impinging in the slab with an angle of incidence $\theta$
is reflected and its amplitude changes depending on the orientation of the external magnetic 
field or intrinsic magnetization of the slab, $\Pi$, which lies in the plane of the slab and 
it is perpendicular to the incidence plane.}
\label{fig1}
\end{figure}

In this work we consider a slab of a NIM with magneto-optical (MO) activity inside a host 
dielectric medium, see Fig.~\ref{fig1}. The MO activity can be due to either an external magnetic 
field or an intrinsic magnetization. We focus here on the analysis of the transverse magneto-optical 
Kerr effect,\cite{Zvezdin1997} which consists in the change of the amplitude of
reflected $p$-polarized light when the magnetic field or the magnetization is in the plane of 
the slab but perpendicular to the plane of incidence, see Fig.~\ref{fig1}. This effect is 
often characterized by the following quantity
\begin{equation}
\mbox{TMOKE} = \frac{R_{pp}(+\Pi) - R_{pp}(-\Pi)}{R_{pp}(+\Pi) + R_{pp}(-\Pi)} ,
\label{tmoke-def}
\end{equation}
which measures the relative change in the reflection probability for $p$-polarized light,
$R_{pp}$, upon reversal of the magnetic field or magnetization, $\Pi$. Notice that the 
maximum of the magnitude of this quantity is equal to 1. The transversal Kerr effect is the 
basis of different sensing applications\cite{Sepulveda2006} and a great effort is being made
to find strategies to enhance this MO effect. Thus, for instance, in the emerging field of magnetoplasmonics
researchers combine ferromagnetic materials with nobel metals in a variety of nanostructures
to use the excitation of plasmons to enhance the TMOKE.\cite{Armelles2013} Although a
lot of progress has been made in this respect in recent years, experimental TMOKE values are 
usually well below $10^{-2}$. In this work we show that the magnitude of TMOKE for a superlens 
with MO activity can reach values very close to 1. In other words, we demonstrate that this system
can behave as an ideal optical filter where the reflected light is suppressed for a given
field or magnetization orientation. Moreover, we show that losses, which are so harmful 
for the imaging capabilities of a superlens, are a necessary ingredient for the occurrence 
of this extraordinary MO effect.

We consider a slab of thickness $d$ of a material that is embedded in a dielectric medium with
a dielectric constant $\varepsilon_0$ and a permeability equal to 1, see Fig.~\ref{fig1}. The 
slab material is characterized by a dielectric constant $\varepsilon = \varepsilon^{\prime} + 
i \varepsilon^{\prime \prime}$ and a permeability $\mu = \mu^{\prime} + i \mu^{\prime \prime}$, 
which can take arbitrary values. Moreover, we assume that this material has MO activity due to 
an in-plane magnetic field or magnetization oriented along the $y$-axis, see Fig.~\ref{fig1}. 
The MO activity in this configuration results in an optical anisotropy that can be described 
by the following dielectric tensor\cite{Zvezdin1997}  
\begin{equation}
\label{perm-tensor}
\hat \varepsilon = \left( \begin{array}{ccc}
\varepsilon & 0 & \varepsilon_{xz} \\ 0 & \varepsilon & 0 \\
- \varepsilon_{xz} & 0 & \varepsilon \end{array} \right) ,
\end{equation}
where $\varepsilon_{xz} = i \alpha m$. Here, $\alpha$ accounts for the magnitude of this 
off-diagonal element and it is simply proportional to the magnitude of the field or the 
magnetization, $\Pi$, whereas $m=\pm 1$ takes into account whether the field or the magnetization 
is parallel ($m=+1$) or antiparallel ($m=-1$) to the $y$-axis. For concreteness, we shall assume 
here that $\alpha$ is a real constant. Notice also that the permeability remains as a diagonal 
tensor in our model.

\begin{figure}[t]
\includegraphics[width=\columnwidth,clip]{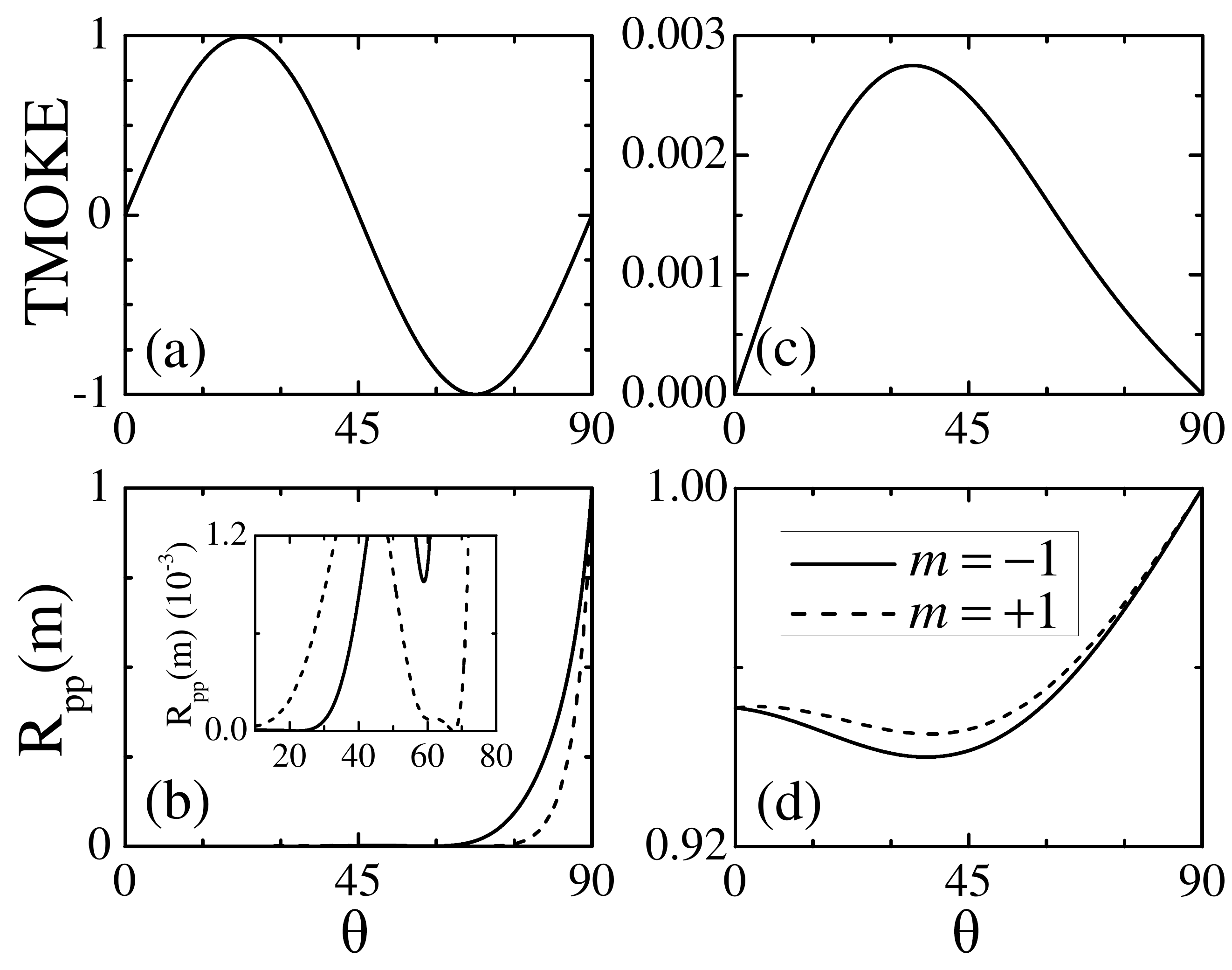}
\caption{(a) TMOKE as a function of the angle of incidence for a slab with $\varepsilon =
-1+0.05i$, $\mu =-1$, and $\alpha = 0.05$, surrounded by air. (b) Reflection probabilities
for $p$-polarized light corresponding to the results of panel (a) for the two orientations
of the field or magnetization. The inset shows a blow up where one can see the suppresion
of these probabilities at certain angles. (c-d) The same as in panels (a) and (b), but for
a permeability equal to $+1$. In all cases the slab thickness, $d$, was chosen to be equal to
the wavelength. While TMOKE does not depend on $d$, the reflection probabilities do.}
\label{fig2}
\end{figure}

As explained above, our goal is to describe the transverse Kerr effect in this system. This requires 
to compute the reflection probability for $p$-polarized waves for a given angle of incidence, $\theta$, 
and for both field/magnetization orientations ($m=\pm1$). This calculation is relatively straightforward 
and it can be done with standard techniques of optics in layered media.\cite{Yeh1988} In our case,
we have carried it out within the scattering matrix formalism\cite{Whittaker1999} adapted to 
anisotropic media.\cite{Caballero2012} The final result for the TMOKE for our configuration
is given by

\begin{widetext}
\begin{equation}
\label{eq-tmoke}
\mbox{TMOKE} = \frac{2\varepsilon_0 \alpha \left[ \varepsilon_0 \gamma 
(\varepsilon^{\prime \prime} \mu^{\prime} + \varepsilon^{\prime} \mu^{\prime \prime})
- 2\varepsilon^{\prime} \varepsilon^{\prime \prime} |\alpha^2 -\varepsilon^2|^2
\cos^2 \theta \right] \sin 2\theta} {\mathcal{D}(\theta)} ,
\end{equation}
where
\begin{eqnarray}
\gamma & = & \alpha^4 + |\varepsilon|^4 + 2\alpha^2 (\varepsilon^{\prime 2} - 
\varepsilon^{\prime \prime 2}), \nonumber \\
\mathcal{D}(\theta) & = & |\alpha^2 - \varepsilon^2|^4 \cos^4 \theta - 2 \varepsilon_0 |\alpha^2
- \varepsilon^2|^2 \cos^2 \theta \left[ \varepsilon_0 (\alpha^2 +
\varepsilon^{\prime \prime 2} - \varepsilon^{\prime 2}) \sin^2 \theta + 
\varepsilon^{\prime} \mu^{\prime} (|\varepsilon|^2 - \alpha^2) +
\varepsilon^{\prime \prime} \mu^{\prime \prime} (|\varepsilon|^2 + \alpha^2) \right] + \nonumber \\
& & \varepsilon^2_0 \gamma \left[ \varepsilon^2_0 \sin^4 \theta + 
\alpha^2 \sin^{2} 2\theta - 2 \varepsilon_0 (\varepsilon^{\prime} \mu^{\prime} -
\varepsilon^{\prime \prime} \mu^{\prime \prime}) \sin^2 \theta +
|\mu \varepsilon|^2 \right] . \nonumber 
\end{eqnarray}
\end{widetext}

Before addressing the case of a superlens, it is worth remarking two general conclusions that
can be drawn from Eq.~(\ref{eq-tmoke}). First, notice that TMOKE is independent of the 
thickness of the slab, although the reflection coefficients $R_{pp}(\pm\Pi)$ are not. 
This is due to the fact that we are considering a symmetric system where the medium of incidence 
and the substrate are made of the same material. More importantly, Eq.~(\ref{eq-tmoke}) tell us 
that TMOKE vanishes if there are no losses in the slab ($\varepsilon^{\prime \prime}, 
\mu^{\prime \prime} = 0$). On the other hand, let us also remind that to observe a 
finite TMOKE one needs to have oblique incidence, and due to the factor $\sin 2\theta$ in 
the numerator of Eq.~(\ref{eq-tmoke}), TMOKE reaches its maximum close to $45^{\rm o}$ in 
conventional structures.

Let us now consider the case of a slab with $\varepsilon^{\prime} = -1$ and $\mu = -1$ surrounded 
by air ($\varepsilon_0 = 1$). In the absence of MO activity, this corresponds to the ideal superlens
considered by Pendry,\cite{Pendry2000} where the transmission is equal to one, irrespective of the
angle of incidence. In Fig.~\ref{fig2}(a) we show the TMOKE as a function of the angle of 
incidence for a case where $\alpha = 0.05$ and $\varepsilon^{\prime \prime} = 0.05$. As one can 
see, the magnitude of TMOKE reaches the value of one at two different angles. More importantly, 
as we show in Fig.~\ref{fig2}(b) these maxima are associated to the blocking of the reflection for 
one of the field or magnetization orientations. Thus, we see that a superlens with MO activity can 
behave as an optical filter exhibiting the highest TMOKE possible. Notice that the values of 
$R_{pp}$ are small for both orientations because we are close to the condition for a perfect lens. 
However, these values can be tuned to some extent by changing the slab thickness without modifying
the TMOKE. For comparison, we show in Fig.~\ref{fig2}(c,d) the corresponding results for a slab where 
the sign of the permeability has been reversed, $\mu = +1$. In this case, the TMOKE has conventional 
values on the order of $10^{-3}$ and there is no filtering effect.

\begin{figure}[t]
\includegraphics[width=0.8\columnwidth,clip]{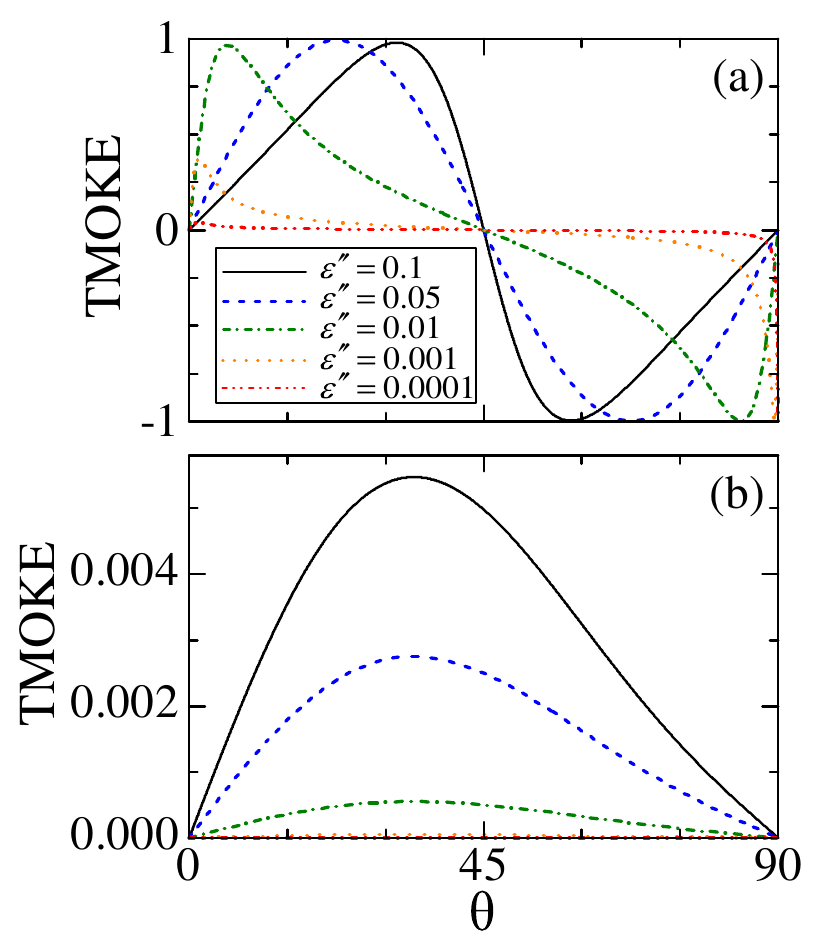}
\caption{(color online) Dependence of the TMOKE with $\varepsilon^{\prime \prime}$ for (a) a superlens
($\varepsilon = -1 + i\varepsilon^{\prime \prime}$ and $\mu=-1$) and (b) a metallic slab
($\varepsilon = -1 + i\varepsilon^{\prime \prime}$ and $\mu=1$). In both cases, $\alpha =
0.05$. Notice that these results are independent of the slab thickness.}
\label{fig3}
\end{figure}

Now, in order to illustrate the role of the losses in this extraordinary TMOKE, we show in 
Fig.~\ref{fig3}(a) TMOKE as a function of $\theta$ for a slab with negative index of refraction, 
but different values of the imaginary part of the dielectric constant, $\varepsilon^{\prime \prime}$, 
while assuming that $\mu$ is real ($\mu^{\prime \prime} = 0$). As one can see, the role of the 
value of $\varepsilon^{\prime \prime}$ is to determine the position of the maxima of $|\mbox{TMOKE}|$, 
which still reach the value of one. Notice that for small values of $\varepsilon^{\prime \prime}$, 
those maxima appear close to $0^{\rm o}$ and $90^{\rm o}$, while they are shifted progressively 
toward $45^{\rm o}$ as the losses increase. Again, this behavior differs dramatically from the 
one observed when $\mu = +1$, see Fig.~\ref{fig3}(b). For the superlens case, the position of 
the maxima $|\mbox{TMOKE}| \approx 1$ can be obtained with the help of Eq.~(\ref{eq-tmoke}). When 
the surrounding medium is air, the positions of the maximum and minimum of TMOKE for 
$\varepsilon = -1 + i\varepsilon^{\prime \prime}$ and $\mu^{\prime \prime} = 0$ are given by 
\begin{equation}
\label{eq-max}
\tan 2\theta = -\frac{m \varepsilon^{\prime \prime}}{\alpha} ,
\end{equation}
where $m=-1$ corresponds to the maximum and $m=+1$ to the minimum of TMOKE, 
respectively. Thus, we see that those positions are simply governed by the ratio between
$\varepsilon^{\prime \prime}$ and $\alpha$, the constant that determines the MO activity
of the slab.

\begin{figure}[t]
\includegraphics[width=0.7\columnwidth,clip]{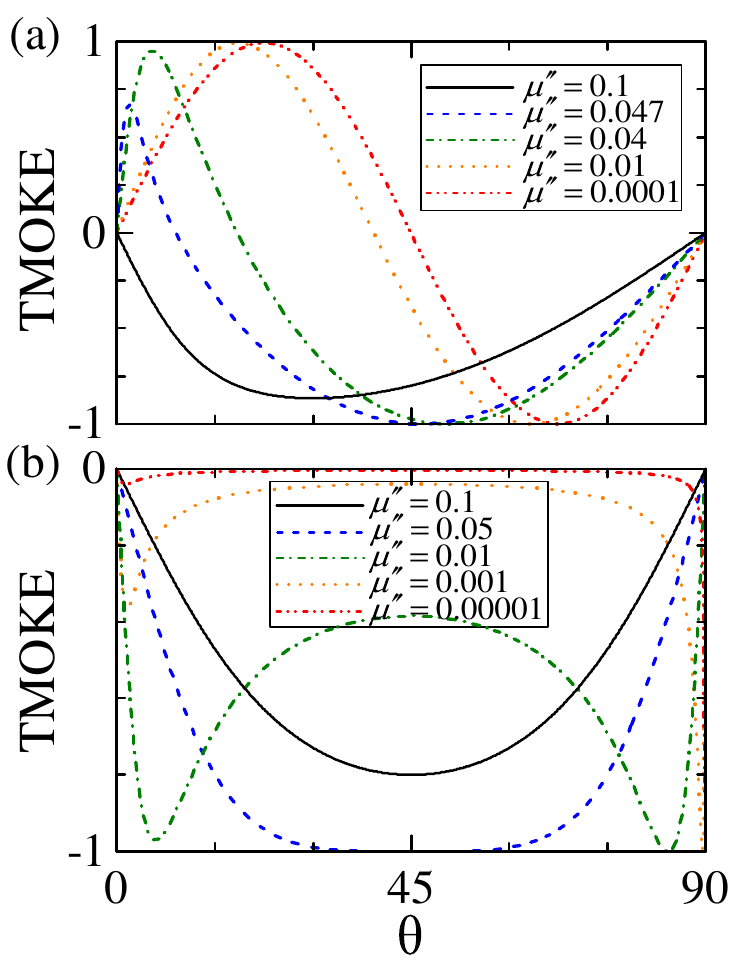}
\caption{(color online) (a) TMOKE as a function of the angle of incidence for a slab
with $\varepsilon= -1 + 0.05i$, $\alpha = 0.05$, and $\mu = -1 + i \mu^{\prime \prime}$. 
The different curves correspond to different values of $\mu^{\prime \prime}$.
(b) The same as in panel (a), but for $\varepsilon = -1$.}
\label{fig4}
\end{figure}

In general, both the permittivity and the permeability can have real and imaginary 
parts.\cite{Xiong2010,Liu2011} For this reason, it is interesting to briefly discuss the role 
of the imaginary part of the permeability, $\mu^{\prime \prime}$. In Fig.~\ref{fig4}(a) we 
show the TMOKE as a function of $\theta$ for the case of Fig.~\ref{fig2}(a), where $\varepsilon= 
-1 + 0.05i$ and $\mu^{\prime} = -1$, for different values of $\mu^{\prime \prime}$. As one can 
see, a finite value of $\mu^{\prime \prime}$ asymmetrizes the angular dependence of the TMOKE, but 
its magnitude can still reach values very close to 1. On the other hand, Fig.~\ref{fig4}(b)
displays the influence of $\mu^{\prime \prime}$ on the angular dependence of the TMOKE for
a case where the permittivity has no imaginary part ($\varepsilon = -1$). In this case,
two features are worth remarking. First, if $\mu^{\prime \prime} = 0$ the TMOKE vanishes,
which is evident from Eq.~(\ref{eq-tmoke}) and confirms that losses are necessary to
observe this MO effect. Second, for finite $\mu^{\prime \prime}$ the magnitude of the
TMOKE can reach again values very close 1 for certain angles of incidence. Thus, we
see that the extraordinary TMOKE may appear for a broad range of values of 
$\varepsilon^{\prime \prime}$ and $\mu^{\prime \prime}$ by simply tuning the angle of incidence.

\begin{figure*}[t]
\includegraphics[width=0.8\textwidth,clip]{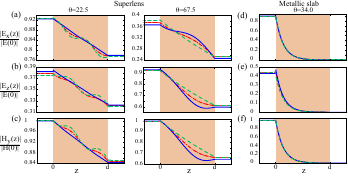}
\caption{(color online) Spatial profiles of non-vanishing field components of a $p$-polarized wave 
propagating through a superlens, panels (a-c), and a metallic slab, panels (d-f). These field distributions
correspond to the maximum (minimum) of TMOKE at $\theta=22.5$ ($\theta=67.5$) in Fig.~\ref{fig2}(a)
for the superlens, and to the maximum occurring at $\theta=34$ in Fig.~\ref{fig2}(c) for the metallic
slab. Dashed and solid curves correspond, respectively, to a magnetized slab with magnetization
parallel ($m=1$) and anti-parallel ($m=-1$) the $y$ axis, and the dash-dotted curves to the
unmagnetized slab ($\alpha=0$).}
\label{fig5}
\end{figure*}

To gain some further insight into the origin of the extraordinary TMOKE of a superlens,
we have analyzed the field distribution inside the slab. In Fig.~\ref{fig5} we show these
distributions for the cases shown in Fig.~\ref{fig2} for those angles at which a maximum
or a minimum of the TMOKE appears. The figure reflects that the effect of the MO element 
in the electromagnetic field profile is substantially larger for the superlens case than
for the conventional metal. In the normal metal case, we can see that the field experiences the 
expected exponential decay inside the metallic layer, with a small perturbation due to the MO
element. In the superlens case the decay within the slab is no longer exponential, but much 
slower. In fact, one can see that the field present some oscillations, except for one of the 
orientations of the magnetic field. Interestingly, it is the ``filtered" component, \emph{i.e.}\
the one that is not reflected, the one with damped oscillations, exhibiting an almost linear 
decay for the three components of the electromagnetic field inside the slab.

\begin{figure}[b]
\includegraphics[width=\columnwidth,clip]{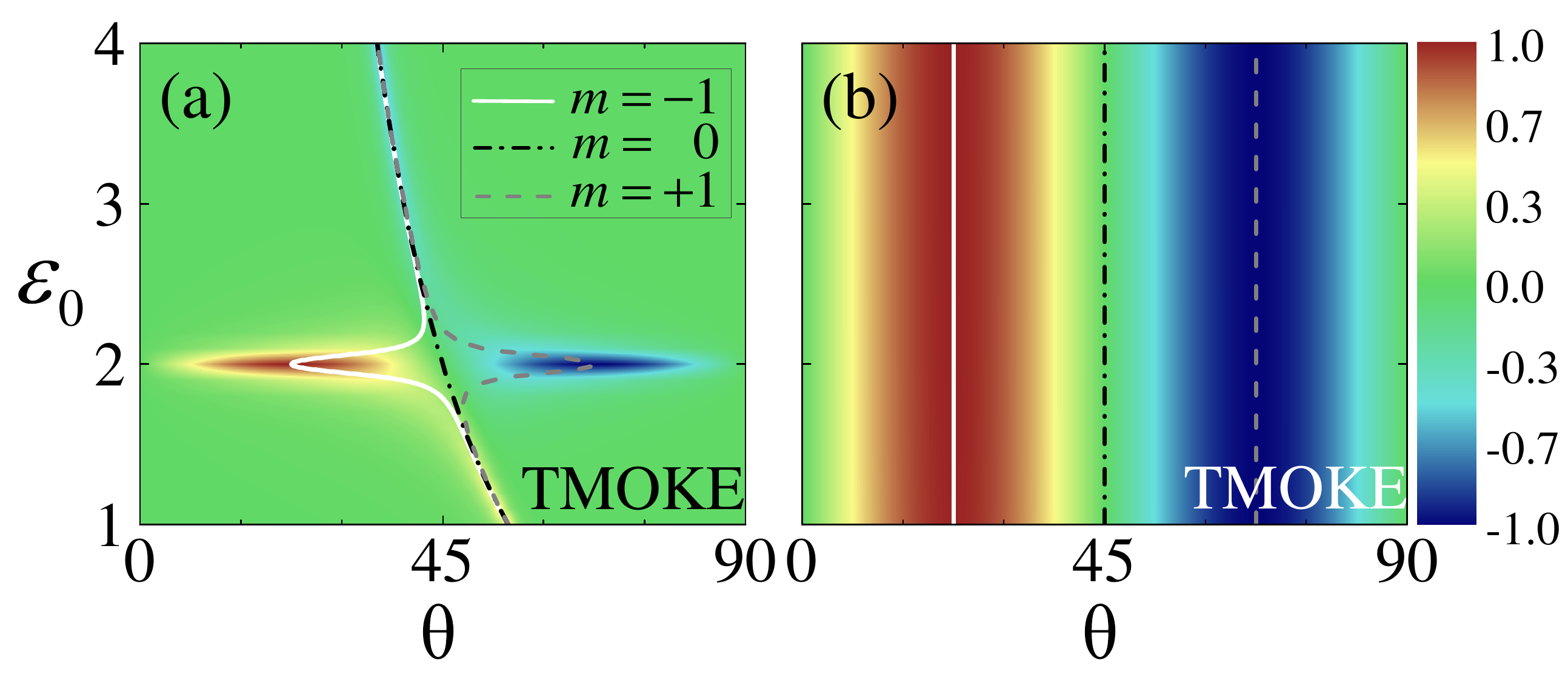}
\caption{(color online) (a) TMOKE as a function of incidence angle and the dielectric constant
$\varepsilon_0$ of the surrounding medium when $\varepsilon^{\prime} = -2$. The other parameters
of the slab are of in Fig.~\ref{fig2}(a). (b) The same as in panel (a), but assuming that
$\varepsilon^{\prime} = - \varepsilon_0$. In both panels the solid and dashed lines 
correspond to the condition of the existence of surface plasmon polaritons, as given by
Eq.~(\ref{eq-SPP}), for the two possible orientations of the magnetization or the magnetic 
field ($m=\pm 1$), while the dotted-dashed corresponds to the condition without MO activity 
($\alpha = 0$).}
\label{fig6}
\end{figure}

As it is evident from Eq.~(\ref{eq-tmoke}), the TMOKE is sensitive to the composition of 
surrounding media. This effect is illustrated in Fig.~\ref{fig6}(a), where we show TMOKE 
as a function of both $\theta$ and $\varepsilon_0$, the dielectric constant of the surrounding 
medium, for $\varepsilon^{\prime} = -2$ and $\mu = -1$. Notice that when $\varepsilon_0 > 
\varepsilon^{\prime}$ the maximum of TMOKE disappears, that is, the reflection coefficient 
$R_{pp}$ is more suppressed at a certain angle when the field or the magnetization is 
parallel to the $y$-axis. However, when $\varepsilon_0 < \varepsilon^{\prime}$ only the maximum 
survives. Notice also that the pattern discussed above with a maximum close to 1 and a minimum 
close to -1 is recovered when the superlens condition $\varepsilon^{\prime} = - \varepsilon_0$ 
is satisified. This is illustrated in Fig.~\ref{fig6}(b) where we show that if this condition 
is fulfilled, one recovers the extraordinary MO effect discussed above. What is so special
about this condition and, more generally, what does determine the extrema of the TMOKE?
It turns out that these extrema are linked to the existence of surface plasmon polaritons
(SPPs) in the interface between the slab and the dielectric. To prove this idea, we 
can make use of the fact that the TMOKE in our symmetric arrangement is independent of 
the slab thickness and consider a semi-infinite slab. In this case, the SPPs supported by 
the interface between the slab and the dielectric are given by the condition
\begin{equation}
\label{eq-SPP}
\eta_0 q_0  + \eta_{xx} q - \eta_{xz} k_{\parallel} = 0 .
\end{equation}
Here, $k_{\parallel} = (\omega /c) \sqrt{\varepsilon_0} \sin\theta$ is the SPP wave vector 
component parallel to the interface and $\omega$ is the radiation frequency. On the other hand, 
$q_0$ and $q$ are the transversal components of the wave vector in the dielectric and in
the slab, respectively, and they are given by $q^2_0 = \varepsilon_0 \omega^2/c^2 - 
k^2_{\parallel}$ and $q^2 = \varepsilon \mu \omega^2/c^2 - k^2_{\parallel}$. Finally, 
$\eta_0 =1/\varepsilon_0$, $\eta_{xx} = \varepsilon /(\varepsilon^2 + \varepsilon_{xz}^2)$,
and $\eta_{xz} = -\varepsilon_{xz} /(\varepsilon^2 + \varepsilon_{xz}^2)$. Equation
(\ref{eq-SPP}) can be solved to obtain the angle of incidence that corresponds to the
existence of the SPPs for both orientations of the magnetic field or the magnetization 
($m=\pm 1$). In Fig.~\ref{fig6}(a,b) we show that the solution of Eq.~(\ref{eq-SPP})
nicely reproduces the position of the extrema of the TMOKE for the whole range of parameters. Thus,
the physical picture that emerges is that at certain angles the reflectivity for a
given orientation of the magnetization or the magnetic field is partially suppressed due 
to the excitation of a SPP, which leads to a maximum or a minimum in the TMOKE. In 
particular, close to the superlens condition ($\varepsilon^{\prime} = - \varepsilon_0$),
the SPPs appear at very different angles of incidence for the two orientations, see 
Fig.~\ref{fig6}(a). Therefore, the extraordinary TMOKE (with a magnitude close to 1) is 
due to the fact that the reflectivity is largerly suppressed \emph{only} for one orientation 
of the magnetic field or the magnetization. It is important to stress that the SPPs 
described here have parallel wave vectors that lie on the left hand side of the light 
lines of both the dielectric and the slab. Thus, these modes are not really confined to 
the interface and they can be excited by a plane wave in the situation described here.   

Let us now briefly discuss possible realizations of a superlens with MO activity. In the
microwave regime, it has been demonstrated, for instance, that metals can be combined with 
insulated ferrites, such as magnetized yttrium iron garnet (YIG), to fabricate metamaterials 
with a tunable negative index of refraction.\cite{Zhao2009} 
Naturally, these hybrid ferromagnetic structures must also exhibit some degree of MO activity. 
Related to this, it has been theoretically suggested that magnetophotonic crystals made of
ferromagnetic nanowires can be good candidates for NIMs and they could exhibit an anomalous 
Faraday effect in virtue of their negative index.\cite{Da2005} On the other hand, although
the realization of a superlens in the optical regime has turned out to be quite challenging,
there have been relatively successful strategies based, for instance, on two-dimensional
metal-insulator-metal (MIM) plasmonic slab waveguides.\cite{Lezec2007} Moreover, it has been
suggested that the practical limitations of these planar MIM geometries can be circumvented by
using coaxial MIM waveguides embedded in a plasmonic crystal.\cite{Burgos2010} Then, the
use of ferromagnetic garnets as insulators in these structures might give rise to systems 
with a simultaneous negative index of refraction and MO activity in the optical regime.

So in summary, we have shown that a superlens made of a slab of a NIM with MO activity could 
exhibit an extraordinary transverse magneto-optical Kerr effect. In particular, we have shown
that irrespective of the slab thickness such a system could act as an ideal optical filter where
losses are necessary, rather than harmful. We have attributed this very peculiar effect to 
the existence of surface plasmon polaritons in the interface between the dielectric and the
slab. Our work illustrates that NIMs may also have a strong impact in the field of magneto-optics 
and we believe that our work may trigger off more theoretical and experimental investigations 
of this interesting interplay.

We thank M.\ Nieto-Vesperinas and J.\ J.\ S\'aenz for enlightening discussions.
E.M.-V.\ was financially supported by the Colombian agency COLCIENCIAS. A.G.-M.\ acknowledges
funding from Spanish Ministry of Economy and Competitiveness through grants 
\textquotedblleft{}FUNCOAT\textquotedblright{} CONSOLIDER CSD2008-00023, and 
\textquotedblleft{}MAPS\textquotedblright{} MAT2011-29194-C02-01, and from Comunidad de Madrid 
through grant \textquotedblleft{}MICROSERES-CM\textquotedblright{}\ S2009/TIC-1476.
J.C.C.\ acknowledges financial support from the Spanish Ministry of Economy and Competitiveness
(Contract No.\ FIS2011-28851-C02-01).

%%%%%%%%%%%%%%%%%%%%%%%%%%%%%%%%%%%%%%%%%%%%%%%%%%%%%%%%%%%%%%%%%%%%%%%%%%%%%%%%%%%%%%%%%%

\end{document}